\documentclass[epj]{svjour}
\usepackage{graphicx}
\begin{document}
\title{Nucleon Form Factors of the Isovector Axial-Vector Current}
\subtitle{Situation of Experiments and Theory}
\author{M.~R.~Schindler \and S.~Scherer 
}                     
%
%
\institute{Institut f\"ur Kernphysik, Johannes
Gutenberg-Universit\"at, D-55099 Mainz, Germany}
\date{Received: date / Revised version: date}
%
\abstract{ The theoretical and experimental status of the isovector axial-vector
current form factors $G_A(q^2)$ and $G_P(q^2)$ of the nucleon is reviewed. We
also describe a new calculation of these form factors in manifestly
Lorentz-invariant chiral perturbation theory (ChPT) with the inclusion of
axial-vector mesons as explicit degrees of freedom.
} 
\maketitle
\section{Introduction}
\label{Introduction}
   The structure of the nucleon is encoded in
several form factors.
   For example, the electromagnetic Dirac and Pauli form factors
$F_1(q^2)$ and $F_2(q^2)$, or equivalently the electric and magnetic Sachs form
factors $G_E(q^2)$ and $G_M(q^2)$, parameterize the matrix elements of the
electromagnetic current operator and are well-known over a wide region of
momentum transfer squared $q^2$.
   For recent reviews on the experimental status of the electromagnetic form
factors see, e.~g., \cite{Gao:2003ag,Friedrich:2003iz,Hyde-Wright:2004gh}.
   In contrast to the electromagnetic case, the three form factors
of the isovector axial-vector current, $G_A(q^2)$, $G_P(q^2)$, and $G_T(q^2)$,
are not as well-known.
   They encode the structure of the matrix elements of the isovector
axial-vector current $A^{\mu,a}(x)$ which, in the SU(2) case, is given by
\begin{equation}
A^{\mu,a}(x)\equiv \bar{q}(x)\gamma^\mu\gamma_5
\frac{\tau^a}{2}q(x), \quad
q=\left(\begin{array}{c}u\\d\end{array}\right),\,  a=1,2,3.
\end{equation}
   $A^{\mu,a}(x)$ is a Hermitian operator that transforms as an
isovector under isospin transformations and as an axial vector under Lorentz
transformations.
   The corresponding matrix element between initial and final nucleon
states is parameterized as
\begin{eqnarray}\label{Def}
\lefteqn{\langle N(p')| A^{\mu,a}(0) |N(p) \rangle}\nonumber\\
&=&\bar{u}(p')\left[\gamma^\mu \gamma_5\, G_A(q^2) +
\frac{q^\mu}{2m_N}\gamma_5 \,G_P(q^2)\right.\nonumber\\
   && \left.
+i\frac{\sigma^{\mu\nu}q_\nu}{2m_N}\gamma_5\,G_T(q^2)
\right]\frac{\tau^a}{2}u(p)\,,
\end{eqnarray}
where $q=p'-p$ and $m_N$ denotes the nucleon mass.
    $G_A(t)$ is called the axial form factor, $G_P(t)$ is the
induced pseudoscalar form factor, and $G_T(t)$ denotes the induced
pseudotensorial form factor.
   For the isospin-symmetric case of equal up and down quark masses, $m_u=m_d$,
the strong interactions are invariant under ${\cal G}$ conjugation, which is a
combination of charge conjugation $\cal C$ and a rotation by $\pi$ about the 2
axis in isospin space (charge symmetry operation),
\begin{equation}
{\cal G}={\cal C}\exp(i\pi I_2).
\end{equation}
   As a consequence the pseudotensorial form factor $G_T(q^2)$ vanishes in
the limit of perfect isospin symmetry.
   So far only upper limits have been established on
the size of $G_T$ \cite{Wilkinson:2000gx}.
   We will neglect $G_T$ in the following discussion.
   From the Hermiticity of the current operator one infers that
$G_A$ and $G_P$ are real functions for space-like momentum transfer squared
($q^2<0$).

   In the Breit system, where $\vec{p}=-\frac{1}{2}\,\vec{q}=-\vec{p}\,'$
and $q_0=0$, the zeroth component of the matrix element of
Eq.~(\ref{Def}) vanishes,
\begin{equation}
\langle N\left({\scriptstyle
\frac{\vec{q}}{2}}\right)|A^{0,a}(0)|N\left({\scriptstyle
    -\frac{\vec{q}}{2}}\right)\rangle=0,
\end{equation}
while the spatial components can be written as\footnote{We adopt the
normalization $\bar u u=2m_N$.}
\begin{eqnarray*}
\lefteqn{\langle N\left({\scriptstyle \frac{\vec{q}}{2}}\right)|
\vec{A}^a(0)
|N\left({\scriptstyle -\frac{\vec{q}}{2}}\right)\rangle=}&&\\
&& \left[2 E\vec{\sigma}_{\perp} G_A(-\vec{q}\,^2)+2m_N\vec{\sigma}_{\|}
D(-\vec{q}\,^2)\right]\frac{\tau^a}{2},
\end{eqnarray*}
   where $E=\sqrt{m_N^2+\vec q^2/4}$, $\vec{\sigma}_\parallel=\vec \sigma\cdot\hat
q\hat q$, $\vec\sigma_\perp= \vec \sigma-\vec{\sigma}_\parallel$, and
\begin{displaymath}
D(q^2)=G_A(q^2)+\frac{q^2}{4m_N^2}G_P(q^2).
\end{displaymath}
   We have thus separated the spatial components of the matrix element into its
transverse and longitudinal parts.
   As we will see below, the longitudinal part vanishes in the chiral limit.

   For recent reviews on the form factors of the isovector axial-vector current see
\cite{Bernard:2001rs,Gorringe:2002xx}.

\section{Axial form factor $G_A(q^2)$}
\label{GA}

   $G_A(q^2)$ is the best known of the isovector axial-vector current form factors.
   Its value at zero momentum transfer squared is defined as the axial-vector
coupling constant $g_A$,
\begin{equation}
    G_A(0)=g_A,
\end{equation}
which has been determined from neutron beta decay to be
\cite{Yao:2006}
\begin{equation}
    G_A(q^2\approx 0)=g_A=1.2695\pm 0.0029.
\end{equation}

   There are two distinct experimental ways of determining the
$q^2$ dependence of $G_A$.
   The first is (quasi)elastic (anti)neutrino scattering (referred to
as simply neutrino scattering in the following).
   For the analysis of experimental data, $G_A(q^2)$ is conventionally
parameterized using a dipole form as
\begin{equation}\label{GAPara}
    G_A(q^2)=\frac{g_A}{(1-\frac{q^2}{M^2_A})^2},
\end{equation}
where $M_A$ is the so-called axial mass.
   The global average for the axial mass extracted from neutrino scattering
experiments given in \cite{Bernard:2001rs} is
\begin{equation}\label{MAv1}
    M_A = (1.026 \pm 0.021)\,\mbox{GeV},
\end{equation}
whereas a recent analysis \cite{Budd:2003wb} finds a slightly smaller value
\begin{equation}\label{MAv2}
    M_A = (1.001 \pm 0.020)\,\mbox{GeV}.
\end{equation}

   The second method makes use of the so-called Adler-Gilman relation \cite{Adler:1966}
which provides a chiral Ward identity establishing a connection between charged
pion electroproduction at threshold and the isovector axial-vector current
evaluated between single-nucleon states (see, e.~g.,
\cite{Scherer:1991cy,Fuchs:2003vw} for more details).
  At threshold (the spatial components of) the center-of-mass transition current
for pion electroproduction can be written in terms of two s-wave amplitudes
$E_{0+}$ and $L_{0+}$
\begin{displaymath}
\left.e\vec{M}\right|_{\rm thr}=\frac{4\pi W}{m_N}\left[i\vec\sigma_\perp
E_{0+}(k^2) +i\vec{\sigma}_\parallel L_{0+}(k^2)\right],
\end{displaymath}
where $W$ is the total center-of-mass energy, $k^2$ is the four momentum transfer
squared of the virtual photon, and $\vec{\sigma}_\parallel=\vec \sigma\cdot\hat
k\hat k$ and $\vec\sigma_\perp= \vec \sigma-\vec{\sigma}_\parallel$.
   The reaction $p(e,e'\pi^+)n$ has been measured at MAMI at an invariant mass
of $W=1125$ MeV (corresponding to a pion center-of-mass momentum of
$|\vec{q}^\ast|=112$ MeV) and photon four-momentum transfers of $-k^2=0.117,
0.195$ and 0.273 GeV$^2$ \cite{Liesenfeld:1999mv}.
   Using an effective-Lagrangian model
an axial mass of
\begin{displaymath}
\bar{M}_A=(1.077\pm 0.039)\,\mbox{GeV}
\end{displaymath}
was extracted, where the bar is used to distinguish the result from the neutrino
scattering value.
    In the meantime, the experiment has been repeated including an
additional value of $-k^2=0.058$ GeV$^2$ \cite{Baumann:2004} and is currently
being analyzed.
    The global average from several pion electroproduction experiments
is given by \cite{Bernard:2001rs}
\begin{equation}\label{MApi}
    \bar{M}_A=(1.068\pm 0.017)\,\mbox{GeV}.
\end{equation}
   It can be seen that the values of Eqs.~(\ref{MAv1}) and (\ref{MAv2})
for the neutrino scattering experiments are smaller than Eq.~(\ref{MApi}) for the
pion electroproduction experiments.
   The discrepancy was explained in heavy baryon chiral
perturbation theory \cite{Bernard:1992ys}.
   It was shown that at order ${\cal O}(q^3)$ pion
loop contributions modify the $k^2$ dependence of the electric dipole amplitude
$E_{0+}$ from which $\bar{M}_A$ is extracted.
   These contributions result in a change of
\begin{equation}\label{deltaMA}
    \Delta M_A = 0.056 \,\mbox{GeV},
\end{equation}
bringing the neutrino scattering and pion electroproduction
results for the axial mass into agreement.

\section{Induced pseudoscalar form factor $G_P(q^2)$}
\label{GP}
   The induced pseudoscalar form factor $G_P(q^2)$ is less known
than the axial form factor $G_A(q^2)$.
   A complete overview over the theoretical and experimental
situation can be found in \cite{Gorringe:2002xx}.

\begin{figure}
\begin{center}
  \includegraphics[width=0.15\textwidth]{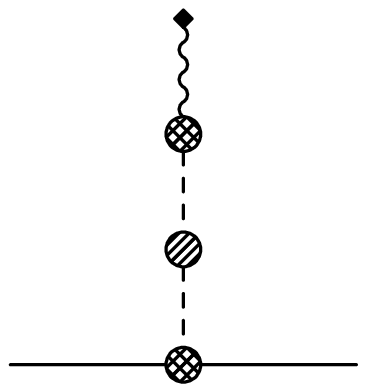}
\caption{Pion pole contribution to $\stackrel{\circ}{G}_P(q^2)$}
\label{fig:pipole}
\end{center}
\end{figure}

   The axial-vector current is divergence-free in the chiral
limit, $\partial_\mu A^{\mu,a}=0$.
   In combination with the Dirac equation one obtains from
Eq.\ (\ref{Def}) in the chiral limit that the equation
\begin{equation}\label{GPChiLim}
    4\stackrel{\circ}{m}^2_N \stackrel{\circ}{G}_A(q^2)+q^2
\stackrel{\circ}{G}_P(q^2)=0
\end{equation}
must hold, where $\stackrel{\circ}{}$ denotes quantities in the chiral limit
(see, e.~g., \cite{Scherer:2002tk} for a detailed discussion).
   Equation (\ref{GPChiLim}) has two possible solutions.
   The first solution is that $\stackrel{\circ}{m}_N=0$ and
$\stackrel{\circ}{G}_P(q^2)=0$.
   However, the nucleon mass in the chiral limit does not vanish.
   The second solution is that $\stackrel{\circ}{G}_A(q^2)$ and
$\stackrel{\circ}{G}_P(q^2)$ are related via
\begin{equation}\label{GPPole}
    \stackrel{\circ}{G}_P(q^2)=
    -\frac{4\stackrel{\circ}{m}^2_N\stackrel{\circ}{G}_A(q^2)}{q^2}.
\end{equation}
   Since $G_A(0)=g_A\approx 1.27$, it follows that
$\stackrel{\circ}{G}_P(q^2)$ has a pole for $q^2\to 0$.
   The behavior of $\stackrel{\circ}{G}_P(q^2)$ for $q^2\to 0$ is
interpreted as stemming from a pion pole contribution (see
Fig.~\ref{fig:pipole}).
   The most general expression for the diagram of
Fig.~\ref{fig:pipole} in the chiral limit is given by
\begin{equation}\label{PoleChiLim}
    - \frac{4 \stackrel{\circ}{m}_N F(q^2) \stackrel{\circ}{g}_{\pi N}(q^2)}{
q^2-\stackrel{\circ}{\tilde\Sigma}(q^2)}\,
\bar{u}(p')\frac{q^\mu}{2\stackrel{\circ}{m}_N}\gamma_5 \frac{\tau^a}{2}u(p),
\end{equation}
where $F(q^2)$ denotes the coupling of the axial source to the pion,
$\stackrel{\circ}{g}_{\pi N}(q^2)$ is the pion nucleon coupling in the chiral
limit and $\stackrel{\circ}{\tilde\Sigma}(q^2)$ stands for the pion self energy
with $\stackrel{\circ}{\tilde \Sigma}(0)=\stackrel{\circ}{\tilde \Sigma'}(0)=0$.
   Comparing with Eq.~(\ref{Def}) one finds the pion pole diagram contribution
\begin{equation}\label{PoleCont}
    -\frac{4\stackrel{\circ}{m}_NF(q^2)\stackrel{\circ}{g}_{\pi N}(q^2)}{
q^2-\stackrel{\circ}{\Sigma'}(q^2)}
\end{equation}
to $\stackrel{\circ}{G}_P(q^2)$.
   The limit $q^2 \to 0$ is given by
\begin{equation}\label{GPqLim}
    \lim_{q^2\to 0} q^2 \stackrel{\circ}{G}_{P,\pi-pole}(q^2)=
-4 \stackrel{\circ}{m}_N F \stackrel{\circ}{g}_{\pi N}
\end{equation}
which by comparison with Eq.~(\ref{GPChiLim}) leads to the famous
Goldberger-Treiman relation
\cite{Goldberger:1958tr,Goldberger:1958vp}
\begin{equation}\label{GTrel}
    \frac{\stackrel{\circ}{g}_A}{F}=\frac{\stackrel{\circ}{g}_{{\pi}N}}{\stackrel{\circ}{m}_N}.
\end{equation}
   While the Goldberger-Treiman relation contains quantities in the
chiral limit, it is interesting to note that it is satisfied to
about 2\% in the real world.

   Information on $G_P(q^2)$ is mainly extracted from muon capture
experiments.
   The induced pseudoscalar coupling $g_P$ is defined as
\begin{equation}\label{gPDef}
   g_P = \frac{m_\mu}{2m_N}\,G_P(q^2=-0.88 m_\mu^2).
\end{equation}
   It is mostly this quantity that has been determined by
experiments, although one pion production experiment measured the
$q^2$ dependence of $G_P(q^2)$ \cite{Choi:1993vt}.
   The weighted world average from ordinary muon capture (OMC),
\begin{equation}\label{OMC}
    \mu^-+p\to \nu_\mu +n,
\end{equation}
is \cite{Bernard:2001rs}
\begin{equation}\label{OMCResult}
    g_P=8.79\pm 1.92\,.
\end{equation}
   Most OMC experiments have used a liquid
hydrogen target, which results in uncertainties due to the
formation of $p\mu p$ molecules.
   For an exact determination of $g_P$ the ortho-para transition
rate has to be known.
   The current MuCap experiment at PSI \cite{MuCap} is using a hydrogen gas
target and thereby avoiding these complications.

   The pseudoscalar coupling $g_P$ can also be determined from radiative muon capture (RMC),
\begin{equation}\label{RMC}
    \mu^-+p\to \nu_\mu +n+\gamma.
\end{equation}
   The value of $g_P$ extracted from these experiments is \cite{Gorringe:2002xx}
\begin{equation}\label{RMCResult}
    g_P = 12.3 \pm 0.9
\end{equation}
and does not agree with the OMC result.
   However, a recent measurement at TRIUMF \cite{Clark:2005as} found a new value for
the ortho-para transition rate in the $p\mu p$ molecule of
\begin{equation}\label{RateNew}
    \Lambda_{op}^{new} = (11.1 \pm 1.7 \pm^{0.9}_{0.6}) \times
    10^4s^{-1},
\end{equation}
which is significantly larger than the previous result
\cite{Bardin:1981cq} of
\begin{equation}\label{RateOld}
    \Lambda_{op}^{old} = (4.1 \pm 1.4) \times 10^4s^{-1}.
\end{equation}
   With the new value the RMC experiment gives \cite{Clark:2005as}
\begin{equation}\label{RMCResultNew}
   g_P = 10.6 \pm 1.1,
\end{equation}
while the average of the liquid hydrogen results is modified to be
\begin{equation}\label{RMCResultLiqNew}
   g_P = 5.6 \pm 4.1.
\end{equation}

   Theoretically, $G_P(q^2)$ has been determined using heavy baryon
chiral perturbation theory \cite{Bernard:1994wn,Fearing:1997dp},
which at order ${\cal O}(q^3)$ gives
\begin{equation}\label{gPTheo}
    g_P = 8.23.
\end{equation}

\section{$G_A(q^2)$ and $G_P(q^2)$ in Lorentz-invariant ChPT}
\label{CHPT}

   With the introduction of renormalization schemes such as infrared
regularization \cite{Becher:1999he} or the extended-on-mass-shell scheme
\cite{Fuchs:2003qc} the calculation of the isovector axial-vector current form
factors in a manifestly Lorentz-invariant formulation of baryon chiral
perturbation theory has been made possible \cite{Ando:2006xy}.

\begin{figure}
\begin{center}
\includegraphics[width=.4\textwidth]{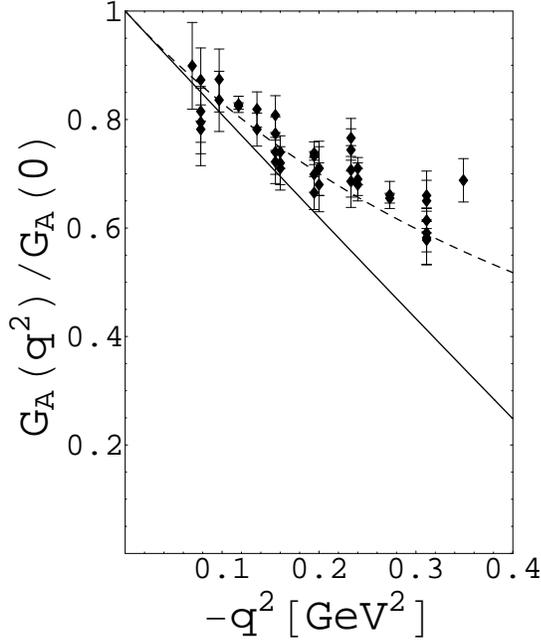}
\caption{$G_A(q^2)$ in chiral perturbation theory at ${\cal
O}(q^4)$ (solid line). The dashed line is a dipole fit. The data compilation is
taken from \cite{Bernard:2001rs}.} \label{fig:GA}
\end{center}
\end{figure}

\begin{figure}
\begin{center}
\includegraphics[width=.4\textwidth]{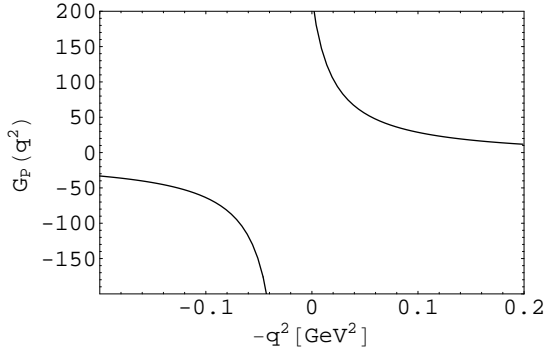}
\caption{$G_P(q^2)$ in chiral perturbation theory at ${\cal
O}(q^4)$.} \label{fig:GP}
\end{center}
\end{figure}

   The results of such a calculation up to and including order ${\cal O}(q^4)$
for $G_A(q^2)$ and $G_P(q^2)$ \cite{Schindler:inprep} are shown in
Fig.~\ref{fig:GA} and Fig.~\ref{fig:GP}, respectively.
   The analytic expression for $G_A(q^2)$ reads
\begin{equation}\label{GAwithout}
    G_A(q^2) = g_A + \frac{1}{6}g_A\langle r^2\rangle_A q^2 +
    \frac{g_A^3}{4F^2}H(q^2),
\end{equation}
where $\langle r^2\rangle_A$ is the mean square axial radius and $H(q^2)$
contains contributions from loop diagrams with $H(0)=H'(0)=0$.
   One can see that the contributions from $H(q^2)$ are small and
the data is only described for very low values of momentum
transfer.
   The calculation at order ${\cal O}(q^4)$ gives
\begin{equation}\label{gPwithout}
    g_P=8.09
\end{equation}
and one can clearly see the pion pole contribution in
Fig.~\ref{fig:GP}.

   The situation for the axial form factor $G_A(q^2)$ can be
compared to the electromagnetic form factors, which in ChPT at
order ${\cal O}(q^4)$ only describe the data for very low momentum
transfers as well.
   In \cite{Schindler:2005ke} a reformulation of infrared
renormalization was used to include vector mesons as explicit
degrees of freedom.
   This resulted in a better description of the data up to about
$-q^2\approx0.4\,\mbox{GeV}^2$.
   Similar to the case of the electromagnetic form factors one can include axial-vector
mesons in the formalism of baryon ChPT \cite{Schindler:inprep}.
   The Lagrangian for the coupling of the axial-vector meson to
pions reads \cite{Ecker:yg}
\begin{equation}\label{Lpi}
    {\cal L}_{{\pi}A}^{(3)} = f_A \mbox{Tr}
    [A_{\mu\nu}F^{\mu\nu}_{-}],
\end{equation}
while the coupling to the nucleon is given by
\begin{equation}\label{LN}
    {\cal L}_{AN}^{(0)} = G_{AN} \bar{\Psi} A^{\mu} \gamma_{\mu}\gamma_5
    \Psi.
\end{equation}
   The contributions of the diagrams containing the axial vector
meson are
\begin{equation}\label{GAAVM}
    G_A^{AVM}(q^2) = - 8 f_A G_{AN} \frac{q^2}{q^2-M_{a_1}^2}
\end{equation}
to the axial form factor and
\begin{equation}\label{GPAVM}
    G_P^{AVM}(q^2) = 32 m_N^2 f_A G_{AN} \frac{1}{q^2-M_{a_1}^2}
\end{equation}
to the induced pseudoscalar form factor, respectively.
   One sees that effectively only one new coupling constant, namely $f_A
G_{AN}$, appears. This coupling constant can be fitted to the data
\cite{Schindler:inprep}.

\section{Summary}
\label{Sum}

   A short overview over the experimental and theoretical situation
of the nucleon form factors of the isovector axial-vector current was given.
   The axial form factor $G_A(q^2)$ has been determined by two
types of experiments, neutrino scattering and pion
electroproduction.
   The results from these two methods agree once pion loop
corrections to the electroproduction amplitude have been taken
into account.
   The situation for the induced pseudoscalar form factor
$G_P(q^2)$ is less clear.
   The results obtained for the induced pseudoscalar coupling $g_P$
from ordinary and radiative muon capture do not agree.
   However, a recent measurement of the ortho-para transition rate in
$p{\mu}p$ molecules results in a significant change of the
previous results for $g_P$.
   A reformulation of the infrared renormalization in baryon ChPT
allows for the inclusion of axial-vector mesons in the calculation of the form
factors, which could result in a better description of the experimental data.
   Only one new low-energy constant effectively appears, which can
be fitted to the data.

\begin{acknowledgement}
This work was supported by the Deutsche Forschungsgemeinschaft
(SFB 443).
\end{acknowledgement}

\end{document}